\documentclass{aa}

\usepackage{graphicx}
\usepackage{txfonts}
\usepackage{natbib}
\usepackage{color}
\definecolor{darkblue}{RGB}{0, 0, 139}
\usepackage[colorlinks=true,backref=true,citecolor=blue,urlcolor=darkblue]{hyperref}

\begin{document}

\title{Analysis of the WN star WR\,102c, its WR nebula,  and the associated 
cluster of massive stars in the Sickle Nebula\thanks{The
scientific results reported in this article are based
on observations obtained during the ESO VLT program 383.D-0323(A)}}

\author{M.\ Steinke
          \inst{1,2}
          \and
          L.\,M.\, Oskinova\inst{1}
          \and
          W.-R. Hamann\inst{1}
          \and
          A. Sander\inst{1}
          \and
          A. Liermann\inst{3}
          \and
          H. Todt\inst{1}}
          \institute{Institute of Physics and Astronomy, University of Potsdam,
              14476 Potsdam, Germany\\
              \email{msteinke@astro.physik.uni-potsdam.de, 
                     lida@astro.physik.uni-potsdam.de}
           \and
         I.\ Physikalisches Institut der
           Universit\"at zu K\"oln, Z\"ulpicher Stra{\ss}e 77, 50937 K\"oln, Germany
           \and
    Leibniz Institute for Astrophysics Potsdam (AIP), An der
Sternwarte 16, 14482 Potsdam, Germany }
   \date{Received 04 November 2015; accepted }

\authorrunning{M.\ Steinke et al.}
\titlerunning{Analysis of WR\,102c and its cluster in the Sickle Nebula }

\abstract
 {The massive Wolf-Rayet type star WR\,102c is located 
near the \object{Quintuplet Cluster},  one of the three massive star clusters in the 
\object{Galactic Centre} region. 
Previous studies indicated that WR\,102c may have a dusty 
circumstellar nebula and is among the main ionising sources of the 
\object{Sickle} Nebula associated with the Quintuplet Cluster.} 
{The goals of our study are to derive the stellar parameters of WR\,102c from the 
analysis 
of its spectrum and to investigate its stellar and nebular environment. }
   {We obtained observations with the ESO VLT integral field 
spectrograph \textsc{SINFONI} in the {\em K}-band, extracted the stellar spectra, and 
analysed them by means of stellar atmosphere models.}
   { Our new analysis supersedes the results previously  reported for WR\,102c. 
   We significantly decrease its bolometric luminosity and hydrogen content.  
   We detect four early OB 
type stars close to \mbox{WR\,102c}. These stars have radial velocities 
similar to that of WR\,102c. 
We suggest that together with WR\,102c these stars belong to a distinct star 
cluster with a total mass of $\sim 1000\,M_\odot$. We identify a new WR nebula 
around WR\,102c in the \textsc{SINFONI} map of the diffuse Br$\gamma$ emission and in the 
{\em HST} Pa$\alpha$ images. The Br$\gamma$ line at different 
locations is not significantly broadened and similar to the width of nebular 
emission elsewhere in the H\,{\sc ii} region around WR\,102c.}
{The massive star WR\,102c located in the Galactic 
Centre region resides in a star cluster containing additional early-type stars. 
The stellar parameters of WR\,102c are typical for hydrogen-free  
WN6 stars. We  identify a   nebula surrounding WR\,102c that has 
a morphology similar to other nebulae around hydrogen-free WR stars, and propose that the 
formation of this nebula is linked to interaction of the fast stellar wind 
with the matter ejected at a previous evolutionary stage of WR\,102c. }  


\maketitle          

\section{Introduction}
The advent of IR observations led to the discovery of many massive 
stars in the 
central part of our Galaxy, mainly congregated in three massive star clusters, 
but 
also scattered in the field \citep[e.g.][]{Figer1999, Hom2003}. The relatively 
isolated 
massive stars in the Galactic Centre (GC) region  have not been extensively observed and are as yet 
poorly understood \citep[among recent results, see 
e.g.][]{dong2015,Habibi2014,Oskinova2013}. 

To study two such relatively isolated Wolf-Rayet (WR) stars, WR\,102c and 
WR\,102ka (known as the \object{Peony star}), we obtained mid-IR spectra using the 
IRS spectrograph on board the  
{\em Spitzer} telescope \citep{houck2004}. These observations lead to the detection of dusty 
circumstellar nebulae  around these objects \citep{Barniske2008}. This was an 
unexpected discovery, since 
it was previously believed that  dust cannot survive in the immediate 
vicinity 
of hot stars of  spectral type WN. However, since then, dusty IR nebulae 
have been discovered around more WN-type stars \citep{gvar2009, gvar2014, burg2013}. 
Even more importantly, 
such dusty nebulae have become an observational ``smoking gun'' in the  search for new WR 
stars
\citep{flagey2011,mau2011,wach2010}.     

\citet{Barniske2008} have performed an analysis of the near-IR {\em K}-band spectrum of
the Peony star by means of the non-LTE stellar atmosphere code PoWR
\citep[e.g.][]{Hamann1998a,Graefener2002}. It has been shown that this object is among the 
most luminous, 
and initially most massive stars in the Galaxy. However, at the  time of their paper 
good 
quality spectra of WR\,102c were not available and only crude qualitative 
conclusions could be made about this source.

\begin{figure*}[t]
  \centering
  \includegraphics[width=17cm]{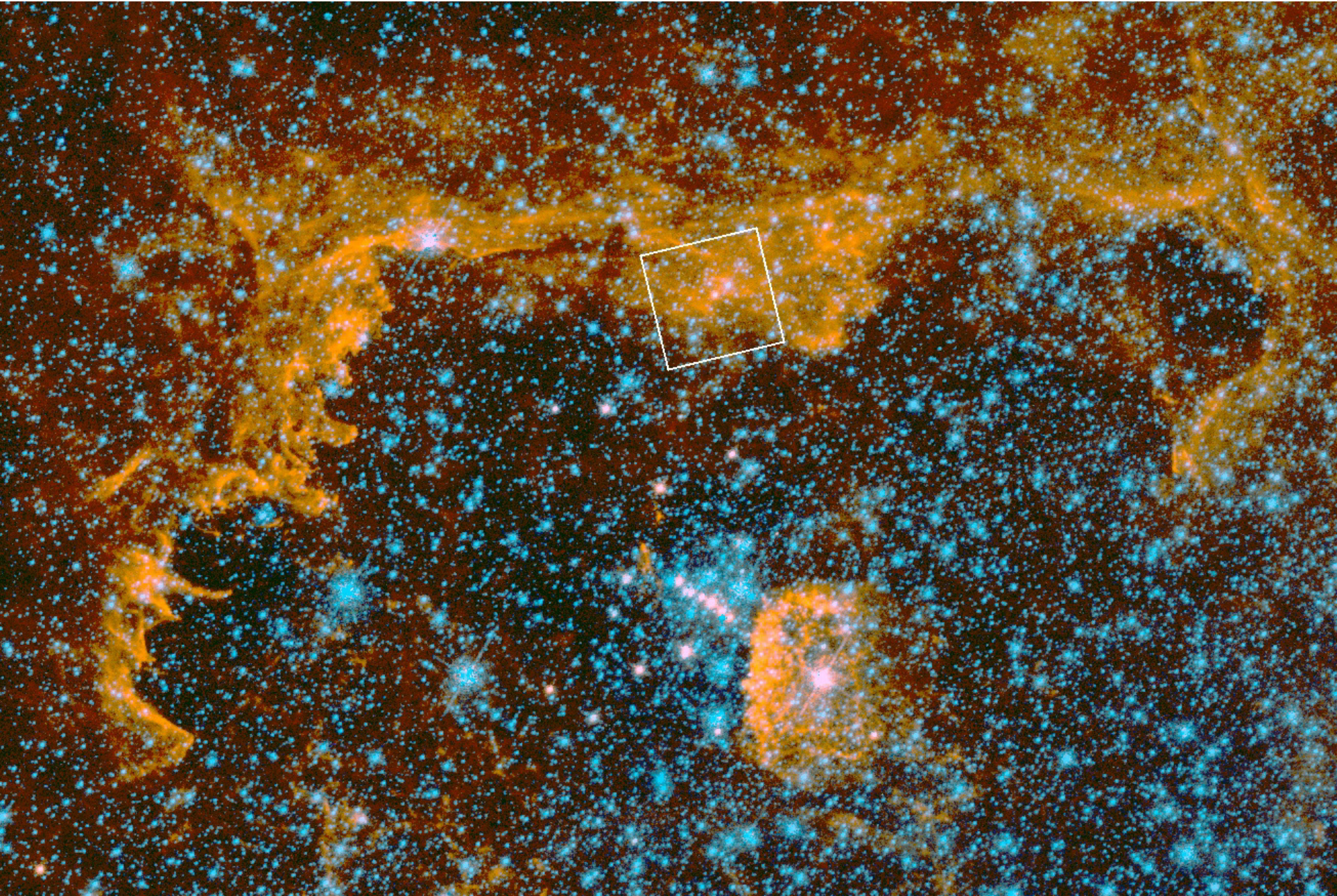}
  \caption[]{\label{fig:sickle} \emph{HST} image of the 
Quintuplet Cluster and the Sickle Nebula. Image size is 
$4\farcm4 \times 2\farcm9$.
The horizontal axis is parallel to Galactic 
longitude. The net
Paschen-$\alpha$ image is in red \citep[see details in][]{Dong2011}, 
the  NICMOS NIC3 F190N image in green, and  the 
NICMOS NIC3 F187N image in blue (continuum and line filter for Pa$\alpha$). 
The white box around WR\,102c 
marks the field of our SINFONI mosaic. This field is $\approx 
1\arcmin$  from the centre of the Quintuplet Cluster, which is seen 
as a group of bright blue stars just below the image centre close to 
the \object{Pistol Nebula} (identified by its characteristic shape).  
The image is based on the data obtained in the Paschen-$\alpha$ 
survey of the Galactic Centre \citep{Wang2010}. }
\end{figure*}

To remedy this situation and to study the Peony star and WR\,102c in more 
detail, we obtained integral field spectra of these stars and their 
nebulae with the 
Spectrograph for Integral Field Spectroscopy in the Near Infrared (\textsc{SINFONI}).
In the previous paper  \citep[herafter Paper I]{Oskinova2013} we analysed 
the Peony star WR\,102ka and its surrounding. In this paper we concentrate on 
the analysis of WR\,102c. 

WR\,102c is located in the neighbourhood of the Quintuplet Cluster, in a large 
H\,{\sc ii} region 
seen in the IR images as extended diffuse emission and  
called the Sickle Nebula (see Fig.\,\ref{fig:sickle}). The star was discovered during a survey 
by \citet{Figer1999}, who classified it as WN6 subtype. 
The goal of our present analysis is to obtain the parameters of WR\,102c
by means of spectral modelling, and to unravel its evolutionary history and  
possible ties with the Quintuplet Cluster. We study the environment of 
WR\,102c by means of integral field spectroscopy to search for other massive 
stars that, together with WR\,102c, may form a massive distinct subcluster. We 
also study the part of the Sickle Nebula that is directly influenced by 
ionising radiation from WR\,102c.  Throughout the paper we adopt distance 
modulus $DM=14.5$\,mag \citep{Reid1993}.    

Special interest in WR\,102c is also motivated by the recent report by 
\citet{Lau2015} who have identified a dusty helix shape filament in the 
vicinity of 
the Quintuplet Cluster. \citeauthor{Lau2015} propose that this dusty filament is a  
precessing, collimated outflow from \mbox{WR\,102c}. 
Furthermore, it has been suggested that WR102c has a gravitationally 
bound compact binary companion with an orbital period $> 800$\,days. 
This hypothesis potentially  makes WR\,102c a very rare and interesting object 
that deserves a thorough study.

The \textsc{SINFONI} observations and data reduction are presented in 
Sect.\,\ref{sec:data}. The circumstellar nebula around WR\,102c is 
discussed in Sect.\,\ref{sec:neb}.
The spectral analysis of WR\,102c is described in Sect.\,\ref{sec:wr102c}. 
The objects detected near WR\,102c, in particular a subcluster of early-type 
stars, are discussed in Sect.\,\ref{sec:fstars}, while the summary and 
the conclusions are drawn in Sect.\,\ref{sec:sum}.

\section{Observations and data reduction}
\label{sec:data}

\begin{figure}[ht!]
  \resizebox{\hsize}{!}{\includegraphics{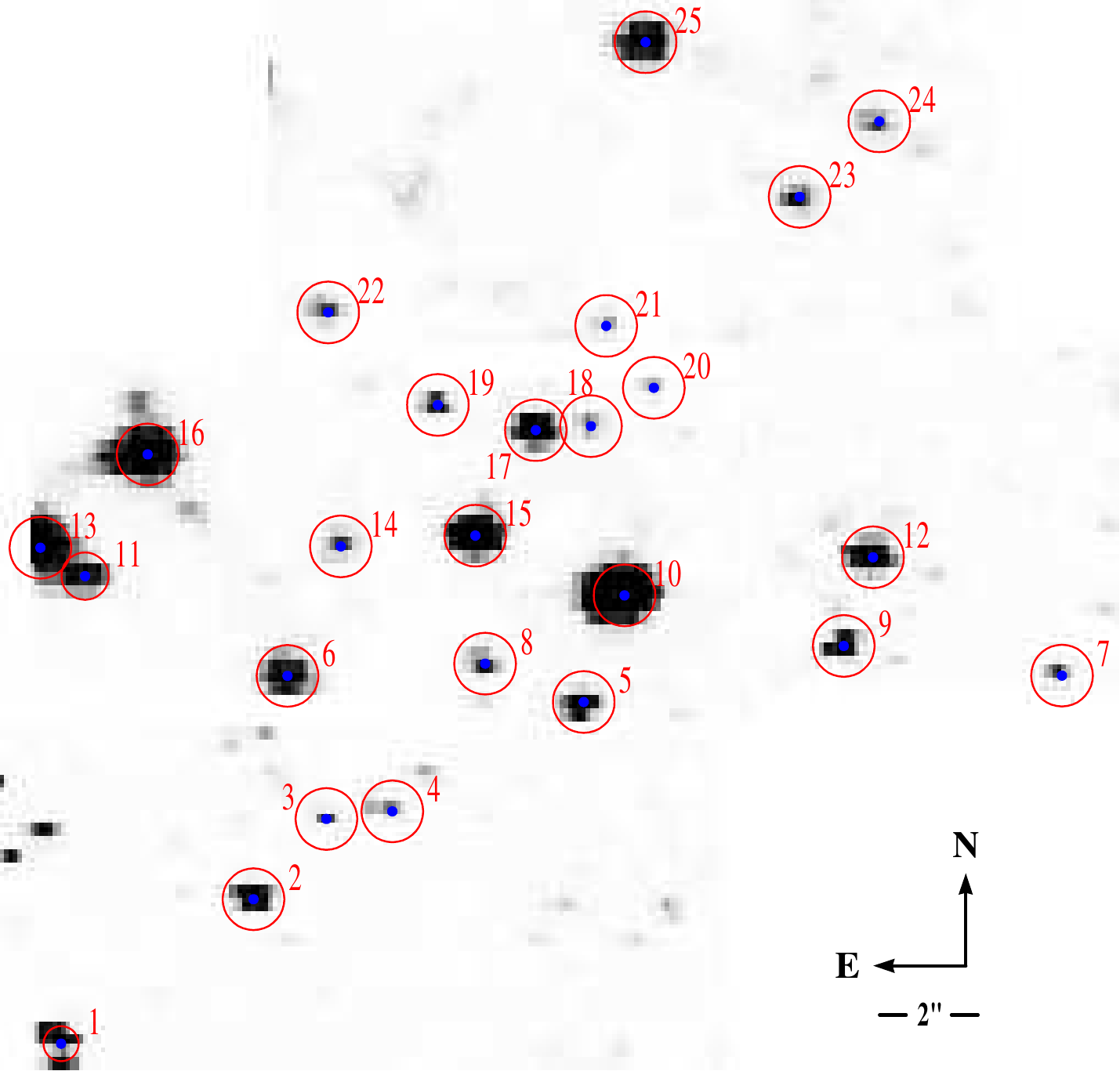}}
\caption{\label{fig:starmap}
Image of the collapsed grand cube (see text for details). The detected sources (open circles) 
are labelled  with their sequential numbers (see\ Table \ref{tab:102c_stars}).
The image size is $23\farcs6 \times 23\farcs1$ ($0.92 \times 0.90$\,pc). 
North is up and east is to the left.}
\end{figure}

The data used in this work were obtained with the ESO VLT UT4
(Yepun) telescope between April 29 and May 19, 2009.
The observations were performed with the integral field spectrograph 
\textsc{SINFONI} 
\citep{Eisenhauer2003,sinf2004} yielding
a three-dimensional data cube with two spatial dimensions
and one spectral dimension.
The  {\em K}-band ($1.95-2.45\,\mu$m) grating with resolving power $R\approx
4000$ was used. The spatial scale was chosen as $0\farcs25$ per pixel. 
The total observation consists of a mosaic of seven pointings  
(observational blocks), covering $\approx 24\arcsec\times
23\arcsec$ ($\approx 0.9$\,pc $\times$ 0.9\,pc)  centred on WR\,102c
(see Fig.\,\ref{fig:starmap}).

The adaptive optics facility could not be used since there is no 
sufficiently bright reference  star in the neighbourhood.
The log of the observations is given in Table\,\ref{tab:observations}. 
ABBA   (science field -- sky -- sky
-- science field) cycles were performed at each pointing. 
Each field was observed with a total exposure time of 300\,s, usually
as the sum of two detector integrations of 150\,s each, except for the
fields 361404 and 361410 where we had to split the integrations into 30\,s 
intervals to avoid saturation of the brightest objects.
To obtain flux calibrated spectra,  standard 
stars were observed at similar 
airmass and in the same mode as the science targets. The seeing was 
between $0\farcs 6$ and $1\farcs 1$, limiting the angular resolution of our 
observations.

\begin{table*}[!ht]
\caption[Observational log file]{Log of observations.}
\label{tab:observations}
\begin{center}
\begin{tabular}{llccrcrr}
\hline
\hline
\noalign{\vspace{1mm}}
 Block  &  Date  & R.A. (J2000)               & Dec. (J2000)       
     &\textsc{sky}\tablefootmark{1} & \multicolumn{1}{c}{Telluric}  & Average \\
  (ID)  &   2009 & $17^{\rm h}46^{\rm m}$ & -28\degr 49\arcmin 
    &  &standard                  & seeing [$\arcsec$]\\
\noalign{\vspace{1mm}}
\hline
361398 & April 29 & 11\fs 10 & -0\farcs 9  & B1 & \object{HIP 83861} & \rule[0mm]{0mm}{3mm} 1.05  \\
361400 & April 29 & 10\fs 56 & -0\farcs 9  & B2 & \object{HIP 83861} & 0.98  \\
361402 & April 29 & 11\fs 47 & 06\farcs 3  & B1 & \object{HIP 83861} & 1.01  \\
361404 & May 17   & 10\fs 93 & 06\farcs 9  & B2 & \object{HIP 88857} & 0.64  \\
361406 & May 17   & 10\fs 37 & 06\farcs 3  & B1 & \object{HIP 85138} & 0.75  \\
361408 & May 19   & 11\fs 51 & 14\farcs 0  & B2 & \object{HIP 86951} & 0.57  \\
361410 & May 19   & 10\fs 97 & 14\farcs 0  & B1 & \object{HIP 84435} & 0.65  \\ \hline
\end{tabular}
\tablefoot{
  \tablefoottext{1}{\textsc{sky} fields B1 and
B2 are centred at 17$^{\rm h}$46$^{\rm m}$18\fs 53, 
$-28$\degr 48\arcmin 5\farcs 7 and 17$^{\rm h}$46$^{\rm m}$15\fs 46, 
$-28$\degr 48\arcmin 54\farcs 2, respectively.}
}
\end{center}
\end{table*}

The data reduction was performed with the \textsc{SINFONI} pipeline version 2.32
({\sc EsoRex} version 3.9.6) using additional cosmic removal from L.A.cosmic
\citep{vanDokkum2001} and self-written tools in  Interactive Data
Language (IDL). 
The procedure was the same as  performed for the Peony star and we refer to section\,2
of Paper\,I for details.

\begin{figure}
  \resizebox{\hsize}{!}{\includegraphics[width=0.9\columnwidth]{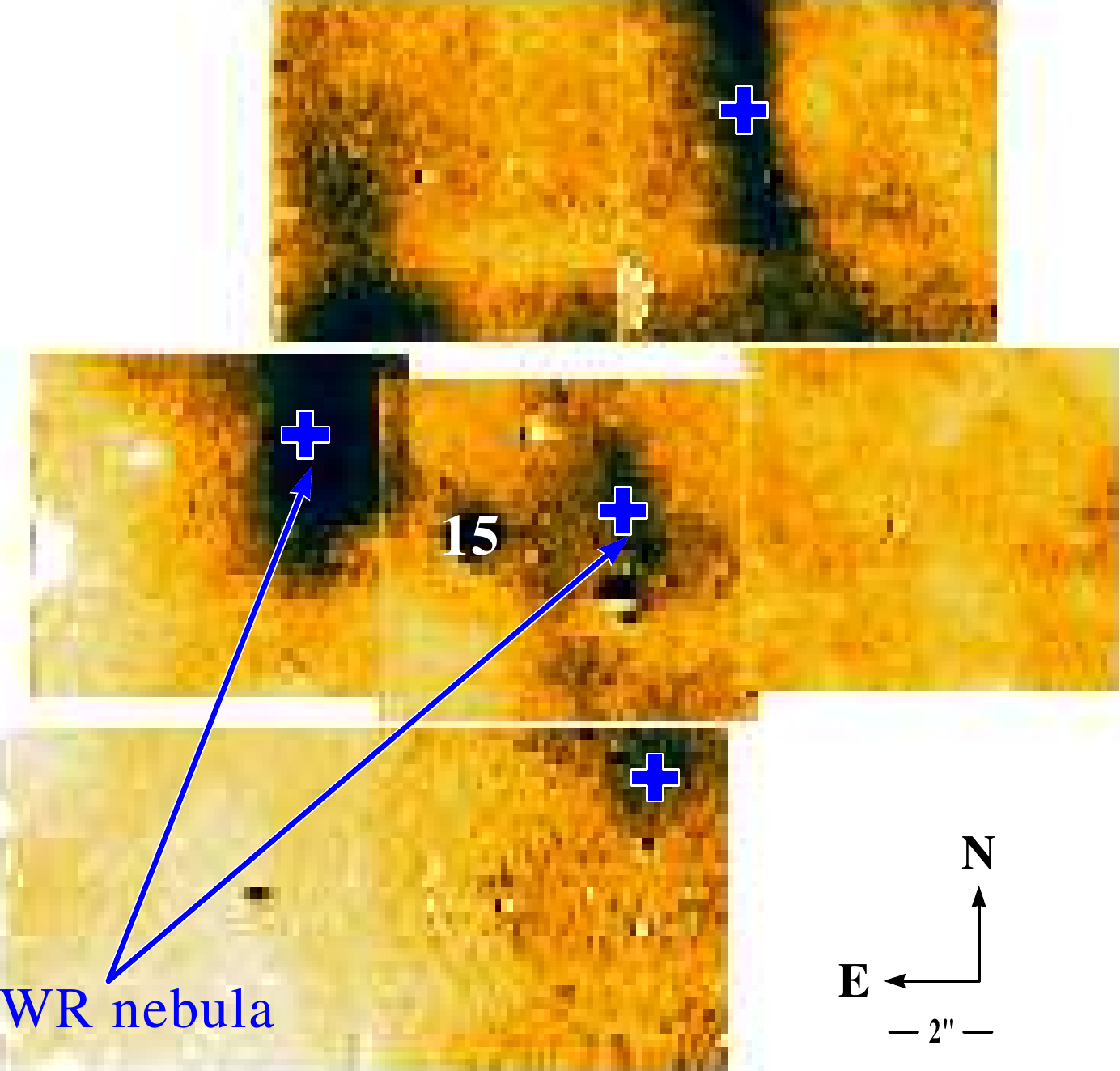}}
  \caption[]{\label{fig:Brgmap}
 Negative image at the wavelength of the Br$\gamma$-line. 
The maximum 
emission (darker areas in the false-colour image) is $\approx 2$\,mJy/arcsec$^2$ while the lowest 
emission (lighter areas) is $\approx 0.3$\,mJy/arcsec$^2$. The blue arrows 
point out the position of the circumstellar \mbox{WR\,102c}-nebula. Source\,15 (WR\,102c) is 
indicated in the centre of the nebula
as are the positions (small crosses) of our extracted spectra\ (cf. Fig.\,\ref{fig:brg}).}
\end{figure}

The flux-calibrated three-dimensional data cubes of the individual fields were 
combined to a grand mosaic cube. This cube was ``collapsed'', i.e.\ summed
over  all wavelengths, to obtain a pass-band image for the purpose 
of point source detection. Our 
observations are sensitive down to an apparent magnitude of $K_{\rm
s}=14.8$\,mag. With the reddening towards WR\,102c (see Sect.\,\ref{sec:wr102c}), the
extinction in the {\emph K}-band amounts to about 3\,mag.  Thus, at the distance
of 8\,kpc, our observations are sensitive to absolute {\emph K}-band magnitudes
$<-2.8$\,mag. 

We securely detect 25 stellar point sources in the observed field (see  
Fig.\,\ref{fig:starmap}). Their 
coordinates, spectral types, and radial velocities are given in 
Table\,\ref{tab:102c_stars}. The \mbox{{\em K}$_{\rm{s}}$}-band magnitudes 
were calculated from the flux-calibrated spectra, 
using the 2MASS filter transmission function \citep{Skrutskie2006}. 
The radial velocities were derived 
from the Doppler shift of prominent lines (except for WR\,102c, which 
has a wind-dominated spectrum), namely the first ${}^{12}$CO band heads for 
cool stars and hydrogen Br$\gamma$ and helium lines for the OB-type stars.
Further details are given in Sects.\,\ref{sec:wr102c} and \ref{sec:fstars}.

\begin{table}[!ht]
\caption[]{Catalogue of stellar sources detected in the observed field (within 
$\approx 0.9$\,pc)
around WR\,102c

}
\label{tab:102c_stars}
\begin{center}
\begin{tabular}{lcccrrr}
\hline
\hline
\noalign{\vspace{1mm}}
No.    & R.A. & Dec.& \mbox{{\em K}$_{\rm{s}}$} & MK & $\varv_{\rm rad}$  \\
  &$17^{\rm h}46^{\rm m}$ & $-28\degr49\arcmin$ &[mag]& type & [km s$^{-1}$]\\
\hline
1 \rule[0mm]{0mm}{4mm}      & 11\fs65 & 17\farcs8 & 12.8 &  M0  & -70  \\
2      & 11\fs38 & 14\farcs9 & 13.0 &  M1  & 95   \\
3      & 11\fs29 & 13\farcs2 & 14.4 &  K1  & -90  \\
4      & 11\fs20 & 13\farcs1 & 14.4 &  K4  & 85   \\
5      & 10\fs94 & 10\farcs8 & 13.4 &  M0  & 55   \\
6      & 11\fs34 & 10\farcs3 & 13.0 &  M0  & 50   \\
7      & 10\fs29 & 10\farcs3 & 14.3 &  K5  & 10   \\
8      & 11\fs07 & 10\farcs1 & 14.0 &  K4  & 205  \\
9      & 10\fs58 & 09\farcs7 & 13.4 &  M0  & 60   \\
10     & 10\fs88 & 08\farcs7 & 10.3 & M5-6 & -45  \\
11     & 11\fs69 & 08\farcs3 & 13.3 &  K5  & 140  \\
12     & 10\fs55 & 07\farcs9 & 12.4 &  M1  & 100  \\
13     & 11\fs67 & 07\farcs7 & 12.0 & B1-2 & 100  \\
14     & 11\fs26 & 07\farcs7 & 14.1 &  B5  & 90   \\
15     & 11\fs08 & 07\farcs5 & 11.6 &  WN6 & 120  \\
16     & 11\fs53 & 05\farcs8 & 12.0 &  K0  & 120  \\
17     & 11\fs00 & 05\farcs3 & 12.6 &  B2: & 70   \\
18     & 10\fs93 & 05\farcs2 & 14.3 &  K3  & 270  \\
19     & 11\fs13 & 04\farcs8 & 13.8 &  K5  & 90   \\
20     & 10\fs84 & 04\farcs5 & 14.8 &  K5  & -250 \\
21     & 10\fs91 & 03\farcs2 & 14.7 &  M0  & 185  \\
22     & 11\fs28 & 02\farcs9 & 14.0 &  M0  & 0    \\
23     & 10\fs64 & 00\farcs6 & 13.8 &  K4  & -65  \\
24     & 10\fs54 & -1\farcs0 & 14.0 &  K2  & -35  \\
25     & 10\fs85 & -2\farcs6 & 12.8 &  O7? & 100  \\
\hline
\end{tabular}
\end{center}  
\end{table}

\section{Circumstellar nebula around WR 102c}
\label{sec:neb}

\begin{figure}
\centering
\resizebox{\hsize}{!}{\includegraphics{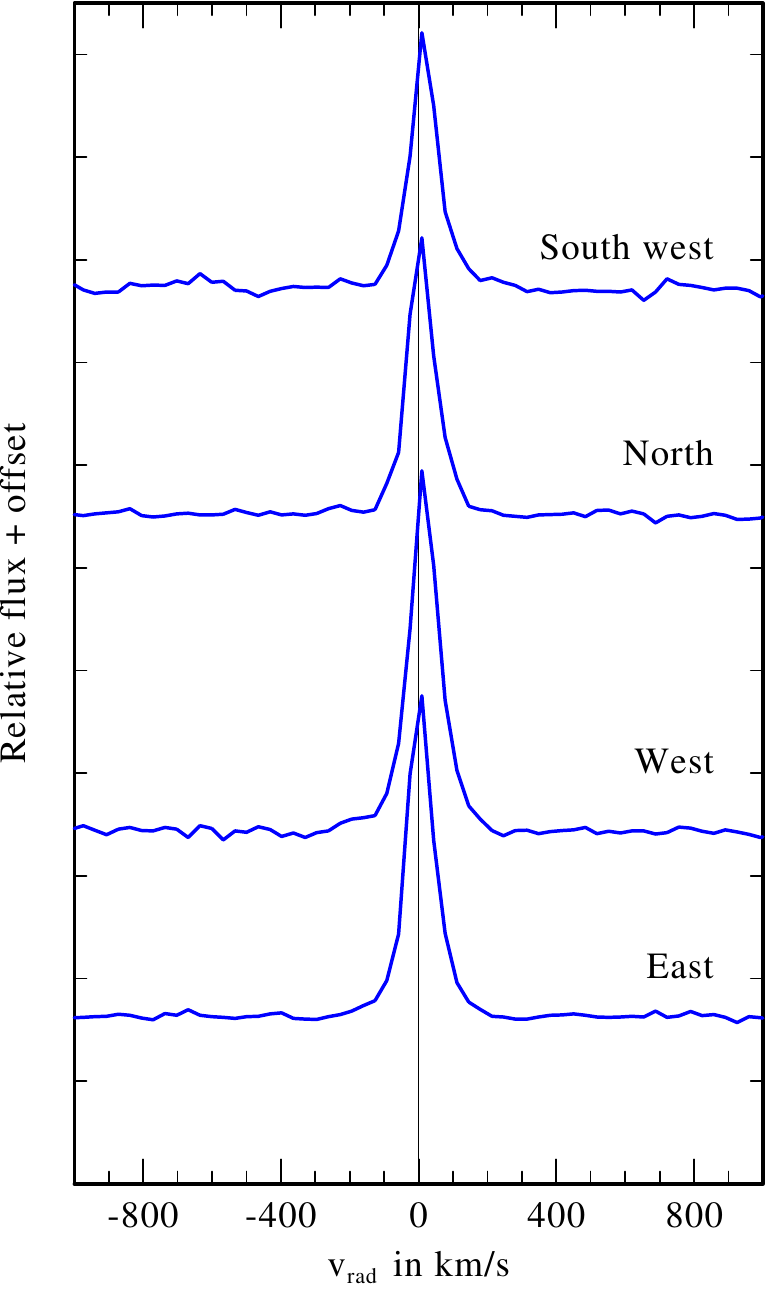}}
  \caption[]{\label{fig:brg}
Observed Br$\gamma$ line in the spectrum of the \mbox{WR\,102c}-nebula at different 
star-free locations. The spectra (as 
labeled) were taken at the following distance and direction from WR\,102c  
east at $4''$ (0.15\,pc), west at $3''$ (0.11\,pc), north at $10''$ 
 (0.38\,pc) south-west at $6''$  
(0.23\,pc), cf.\ Fig.\,\ref{fig:Brgmap} for the positions.}
\end{figure}

From the analysis of mid-IR observations of WR\,102c with the {\em Spitzer} and 
MSX telescopes, \citet{Barniske2008} detected  a dusty circumstellar nebula 
heated by 
the intense radiation of the WR star. However, the spatial 
resolution of these instruments was not sufficient to directly image this 
nebula. Our near-IR \textsc{SINFONI} data have a much better resolution. Hence, to study 
the nebular emission we created a narrow band image in the Br$\gamma$ line. 
A bipolar or ring structure centred on WR\,102c 
(source 15) is clearly seen in this image (see Fig.\,\ref{fig:Brgmap}).   
The radius of the nebula is $\approx 4^{\prime\prime}$ ($0.15$\,pc). 
As the field is quite crowded with stars, it is difficult 
to conclude whether the nebula is circular or bipolar.  To measure its 
expansion velocity, the spectra were extracted at different locations. 
As can be seen in Fig.\,\ref{fig:brg}, the nebular Br$\gamma$ lines are neither 
shifted nor broadened by more than $\approx 80$\,km\,s$^{-1}$ 
(\textsc{SINFONI}'s spectral resolution). This upper limit on the expansion 
velocity indicates 
that the nebula is not  expanding quickly, opposite to what would be 
expected if the observed bipolar structure were the result of a fast 
collimated outflow.

\begin{figure*}
\sidecaption
\includegraphics[width=12cm]{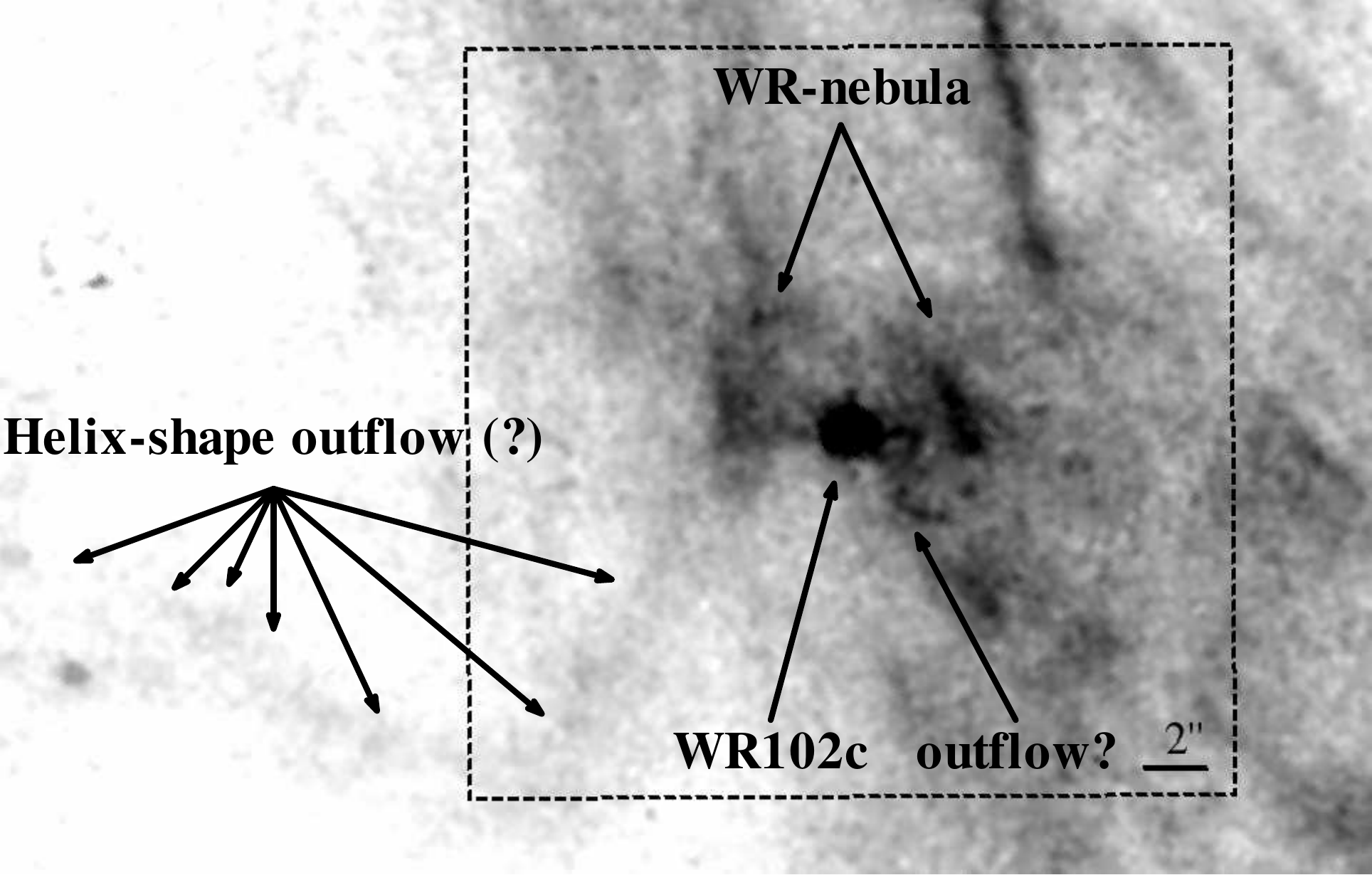}
  \caption{\label{fig:ring}
 {\em HST} Paschen-$\alpha$ image \citep[][]{Dong2011} of the 
ring nebula around \mbox{WR\,102c}. The legends and the arrows indicate
(i) the helix-like nebulosity detected by \citet{Lau2015}; 
(ii) the ring WR nebula as seen 
in the SINFONI Br$\gamma$ image in Fig.\,\ref{fig:Brgmap}; 
(iii) a small structure that  resembles a helix-like outflow.    
The dashed rectangle corresponds to our SINFONI mosaic field. 
The orientation is the same as in Figs.\,\ref{fig:starmap} 
and \ref{fig:Brgmap}, i.e.\ north is up and east is to the left.   
The image is a zoom into the Paschen-$\alpha$ survey of the 
Galactic Centre \citep{Wang2010}.}
\end{figure*}

Nebulae with various morphologies are often 
observed around WR stars. The mechanisms of nebula formation are linked to  
stellar evolution: when a massive star leaves the main sequence, it loses 
a large amount of matter. In the luminous blue variable (LBV) or a 
red supergiant (RSG) phase the stellar wind is dense and slow, while at a 
later evolutionary stages (WN) the wind is fast. The interaction between the 
slow and 
fast moving wind creates shells that are ionised and heated by the
stellar radiation \citep[e.g.][]{gs1995,Freyer2006,vMarle2007,Toala2011}. A 
recent review on WR nebulae is given in \citet{Toala2015} and 
references therein.  

To explain the nature of the nebula around WR\,102c
seen in the Br$\gamma$ image we considered the following check-list. The 
nebula is surrounding a WR-type star. It has a small expansion velocity 
similar to other WR nebulae \citep[e.g.][]{Marston1999}. The nebula is  
clearly seen from its hydrogen emission (in this case Pa$\alpha$ and Br$\gamma$) 
and in He\,{\sc i}
maps of diffuse gas, analogously to other WR nebulae. 
Significantly, WR nebulae have morphologies that correlate with the spectral 
type of their central stars. The ring or bubble-like morphologies ($\emph 
B$-type), sometimes bipolar, are found around WNE stars with fast winds, just 
as in the case of \mbox{WR\,102c}.  These characteristics lead us  to 
conclude that the nebula around WR\,102c is a typical WR nebula. 

However, there are some important aspects that distinguish the WR\,102c nebula. It 
is small,   only 0.15\,pc, while the typical size of WR nebulae is a 
few parsec. The small size of this nebula can, perhaps, be attributed to the 
very dense and warm environment in the vicinity of the Quintuplet 
Cluster. Another unusual aspect is the presence of a 
helix-like tail likely associated with WR\,102c as found by \citet{Lau2015}.  

The \mbox{WR\,102c}-nebula and helix tail can also be clearly recognised in the 
{\em HST} Pa$\alpha$ image (Fig.\,\ref{fig:ring}). Moreover, it appears that 
another helix-like outflow is seen {\em within} the ring-nebula, as indicated 
in  Fig.\,\ref{fig:ring}. However,  
the Pa$\alpha$ images should be interpreted with caution. 
These were heavily processed, as explained in depth 
in \citet{Dong2011}. The \textsc{SINFONI} images do not show obvious helix-like 
structures associated with \mbox{WR\,102c}.  

\citet{Vink2011} have found a significant correlation between fast 
rotating WR stars and the subset of WR stars with ejecta nebulae. They note that these 
objects have only recently transitioned from a previous RSG or 
LBV phase and thus have not yet significantly undergone spin-down 
which will be caused by their intensive mass 
loss due to the conservation of momentum.
This subset of fast rotating WR stars are the candidate $\gamma$-ray 
burst progenitors.
\citet{Graf2012} have identified an incidence rate of $\sim 23$\,\%\ for WR stars 
in the Galaxy that have possible ejecta nebulae and rotate faster than the 
average WR stars. 
They note that even early-type hydrogen-free WR stars, such as WR\,6 
(WN4) and WR\,136 (WN6) can be fast rotators, perhaps evolving through 
binary channels.
Hence, the presence of a compact bipolar nebula and a 
helix-like outflow from WR\,102c may indicate that this object has only recently 
transitioned to the WR evolutionary stage, and is still a relatively fast 
rotator.

\section{The WN6 star WR 102c}
\label{sec:wr102c}

WR\,102c was discussed in \citet{Barniske2008} who confirmed its spectral 
type as WN6 based on the inspection of the low-resolution spectrum 
shown by \citet{Figer1999}. In the present paper, we now analyse the new 
high-quality \textsc{SINFONI} spectrum of WR\,102c in detail  
largely revising the estimates made in \citet{Barniske2008}.

The WR\,102c spectrum (Fig.\,\ref{fig:WR102c_spectrum}) was analysed 
by means of PoWR model 
atmospheres\footnote{\mbox{\url{http://www.astro.physik.uni-potsdam.de/PoWR/}}} 
\citep{Ham2004,Todt2015}. The PoWR code
has been used extensively to analyse the  spectra of massive stars in
the IR as well in the ultraviolet and optical range
\citep[e.g.][]{osk2007, lho2010, osk2011,Sander2012,Hainich2014}. The PoWR 
code solves the non-LTE radiative transfer in a spherically expanding atmosphere
simultaneously with the statistical equilibrium equations and accounts
at  the same time for energy conservation. Complex model atoms with
hundreds  of levels and thousands of transitions are taken into
account. The computations for the present paper include detailed model
atoms for hydrogen, helium, carbon, oxygen, nitrogen, and silicon. Iron
and iron-group elements with millions of lines are included through the
concept of super-levels \citep{Graefener2002}. The extensive inclusion of
the iron group elements is important because of their blanketing effect
on the  atmospheric structure.

The stellar radius $R_\ast$, which is the inner boundary of our model 
atmosphere, corresponds by definition to a Rosseland continuum optical depth of 
20. The 
stellar temperature $T_\ast$ is defined by the luminosity and the stellar 
radius $R_\ast$ via the Stefan-Boltzmann law; i.e.\ $T_\ast$ denotes the 
effective temperature referring to the radius $R_\ast$. For the wind velocity 
we adopt the usual $\beta$-law with $\beta = 1$ \citep[e.g.][]{Sander2015}.
Each stellar atmosphere model is defined by its effective temperature,
surface gravity, luminosity, mass-loss rate, wind velocity, and chemical
composition. 
The gravity determines the density structure of the
stellar atmosphere below and close to the sonic point.
If lines from the quasi-hydrostatic layers are visible in the spectrum, the
analysis allows  the gravity and thus the stellar mass to be derived from the
pressure-broadened absorption line profiles \citep{Sander2015}.

For a given chemical composition and stellar temperature $T_\ast$,
synthetic spectra from WR model atmospheres of different mass-loss
rates, stellar radii, and terminal wind velocities yield almost the same
emission-line equivalent widths if their \emph{transformed
radii}
$
R_{\rm t} \propto R_\ast \left[\varv_\infty D^{-1/2} \dot{M}^{-1} \right]^{2/3}
$    \citep{Schmutz1989} agree.   
To account for small-scale wind inhomogeneities, we adopt a clumping contrast 
of $D=4$ as a typical value for WN
stars \citep{Hamann1998}. We note that $R_{\rm t}$ is inversely 
correlated with the mass-loss rate, i.e.\ the smaller the transformed 
radius, the higher  the density in the stellar wind.

\subsection{Temperature of WR 102c}

\begin{figure*}
  \centering
\includegraphics[width=17cm]{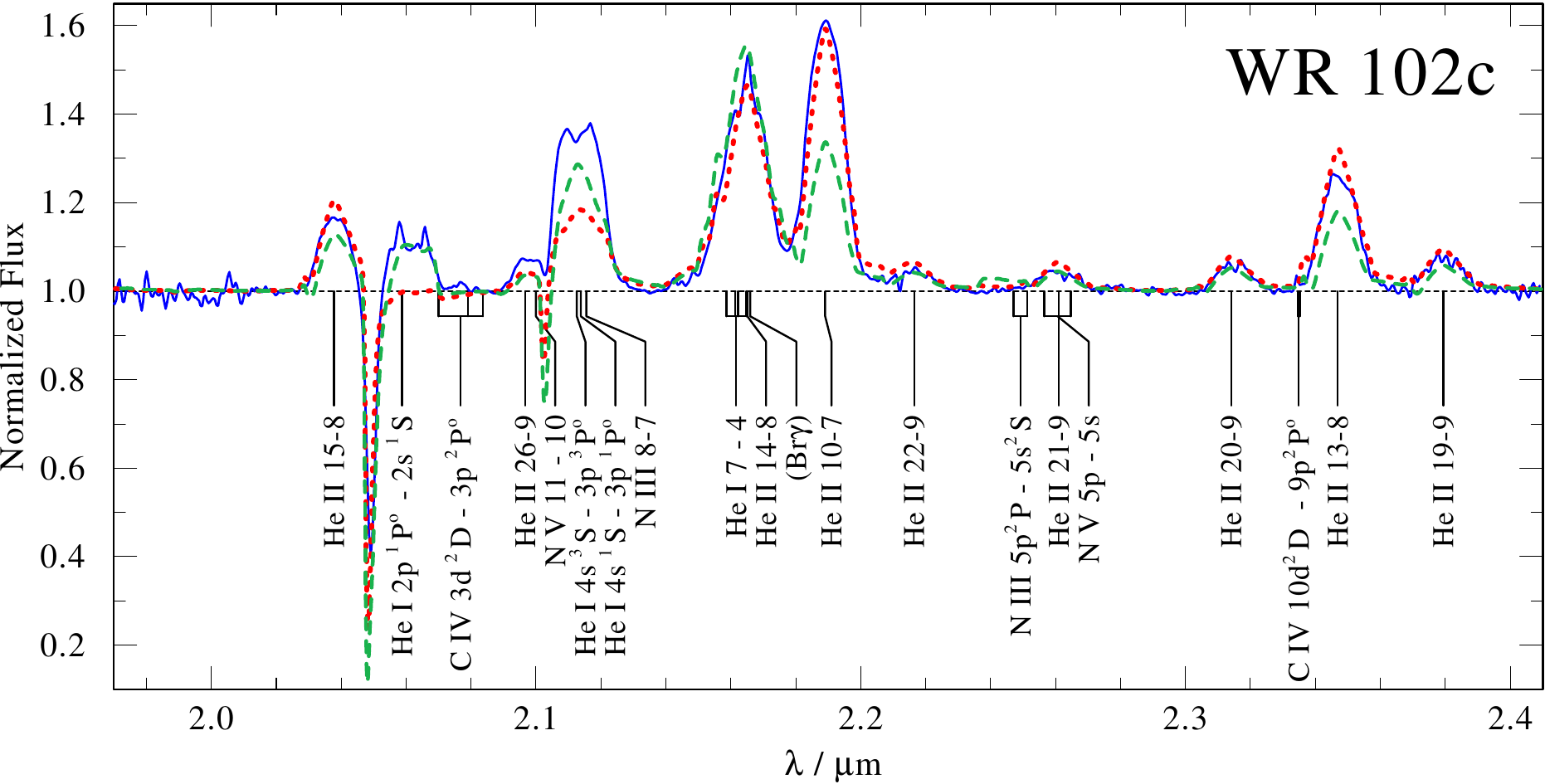} 
\caption{\label{fig:WR102c_spectrum} 
 Normalised {\em K}-band spectrum of WR\,102c (solid blue), compared to two PoWR models:
$T_\ast=75.0\,$kK, $\log L/L_\odot=5.57$, $\log\dot{M}=-4.17$ (red dotted line);
$T_\ast=66.0\,$kK, $\log L/L_\odot=5.51$, $\log\dot{M}=-4.08$ (green dashed line).
The prominent spectral features are identified.
}
\end{figure*}

Figure \ref{fig:WR102c_spectrum} shows the normalised spectrum of WR\,102c
compared to synthetic spectra. Unfortunately, it is impossible to find
a model that simultaneously fits all \ion{He}{i} and \ion{He}{ii} emission
lines. The model for $T_\ast = 75\,$kK reproduces the \ion{He}{ii} emission
lines perfectly, but the \ion{He}{i} lines at $2.06\,\mu$m and  $2.11\,\mu$m 
are severely underestimated (red dotted line in Fig.\,\ref{fig:WR102c_spectrum}).
The alternative model drawn as a green dashed line for a temperature of 
$T_\ast = 66\,$kK reproduces the \ion{He}{i} line at $2.06\,\mu$m perfectly,
but underestimates some of the \ion{He}{ii} lines. The line at
$2.11\,\mu$m is actually not only composed of \ion{He}{i}, but is also blended
with \ion{N}{iii}. The corresponding transitions are not fully 
included in our atomic data, which might explain the mismatch.

Hence, the stellar temperature remains uncertain in the range 
$T_\ast = 66\,$kK to $75\,$kK. Moreover, if the star 
is a fast rotator it could lead to variations of the
wind properties between poles and equator.

It is not possible to establish the rotation rates
of WR stars from spectroscopic analysis of the near-infrared only. 
For example,  our test calculations show that even the unrealistically high 
photospheric rotation velocities up to $\varv_{\rm rot} \sin i = 
1000\,$km\,s$^{-1} > \varv_{\rm crit}$
would not leave any noticeable imprint on the {\em K}-band spectrum as 
long as the wind does not corotate \citep[see discussion in][]{Shenar2014}.
We do not observe signs of such extreme wind corotation in WR\,102c.

Fortunately, the impact of the $T_\ast$ uncertainty on the implied mass-loss 
rate and luminosity is relatively small (see\ Table \ref{tab:stellarParameters102c}). 
In particular,  the effect on $T_{2/3}$, i.e.\ the effective temperature referring to the
radius of Rosseland optical depth $\tau_{\rm Ross} = 2/3$, is hardly affected.
This invariance reflects the parameter degeneracy of very dense stellar winds
as discussed in e.g.\ \citet{Hamann2006}.

Another speculative explanation
for the inconsistent spectral fit could be the presence of a cooler, unresolved companion.
A useful empirical diagnostic of WR-binaries is provided by X-ray observations. 
While single WR stars are relatively faint X-ray 
sources \citep{Ignace2000, osk2003}, the WR-type binaries usually display 
colliding wind phenomena and are bright, detectable X-ray sources 
even in the GC region \citep{law2004,osk2005}. WR\,102c was not  
detected in deep X-ray surveys of the Quintuplet Cluster \citep{Wang2006}. 
This non-detection provides evidence against the binarity hypothesis.  

\subsection{Abundances}

\citet{Barniske2008} constrained the hydrogen abundance in WR\,102c as being 
below 20\%\ by mass. The new data allow  this limit to be adjusted to $< 5$\%. If 
the hydrogen content were higher, it would contribute significantly to 
the emission at $\lambda=2.16\,\mu$m (\ion{He}{ii} + Br$\gamma$). Thus, our new 
analysis confirms that WR\,102c is a virtually hydrogen-free WNE star. 

In the observed spectrum, all clearly detectable metal lines (of nitrogen and carbon) are 
blended with helium lines (e.g.\ \ion{N}{v} at $\lambda2.1\,\mu$m
or C\,{\sc iv} at $\lambda2.08\,\mu$m). For our models we adopted a mass fraction 
of $1.5\,\%$ for nitrogen, which is typical for Galactic WN stars.  
The carbon abundance is constrained to $\ll 1\%$, otherwise carbon emission lines 
would be visible (e.g.\ C\,{\sc iv} at  $\lambda=2.07-2.08\,\mu$m). Hence, we
set the carbon abundance to a mass fraction of $10^{-4}$. 

\subsection{Bolometric luminosity of WR 102c}

\begin{table}[ht!]
\begin{center}
\caption[]{Stellar parameters of WR\,102c} 
\label{tab:stellarParameters102c}
\begin{tabular}{lr}
\hline
\hline
\noalign{\vspace{1mm}}
Spectral type  \rule[0mm]{0mm}{3mm}       &WN6      \\
$\log L/L_\odot$ & $5.5\ldots5.7$       \\
$T_\ast$ [kK]    & $66\ldots 75$  \\
$T_{2/3}$ [kK]   & $33\ldots 37$  \\
$\log\dot{M}\, [M_\odot\,{\rm yr}^{-1}]$
                  & $-4.17 \ldots -4.09$  \\
$\log\,\Phi_{{\rm Ly}}$ [s$^{-1}$] & $49.0 \ldots 49.3$ \\ 
$\varv_{\infty}$ [km s$^{-1}$]
                 & 1600     \\
$M_\mathrm{ini}/M_\odot$
                 & $\sim 40$       \\
                 
$\varv_{\rm{rad}}$ [km s$^{-1}$]
                 & $120 \pm 40$   \\
$A_{K}$ [mag]      & 2.8      \\
\hline
\end{tabular}
\end{center} 
\end{table}

In their discovery paper, \citet{Figer1999} assigned a \mbox{{\em K}$_{\rm{s}}$}-band 
magnitude of 11.6\,mag to \mbox{WR\,102c}. This value was revised by 
\citet{Barniske2008} 
who 
erroneously used the \mbox{{\em K}$_{\rm{s}}$}-band magnitude of a nearby bright 
star (source 10 in 
Fig.\,\ref{fig:starmap}) from the 2MASS 
and the {\em Spitzer} IRAC point source catalogues and adopted $K_{\rm s}=9.3\,$mag 
for \mbox{WR\,102c}. This resulted in an overestimated stellar bolometric luminosity. 
The new \textsc{SINFONI} data allow us to unambiguously identify WR\,102c in the near-IR 
image (source 15 in Fig.\,\ref{fig:starmap}) with \mbox{{\em 
K}$_{\rm{s}}$}=11.6\,mag, in 
agreement with the original report. 

\begin{figure*}[ht!]
  \sidecaption
\includegraphics[width=12cm]{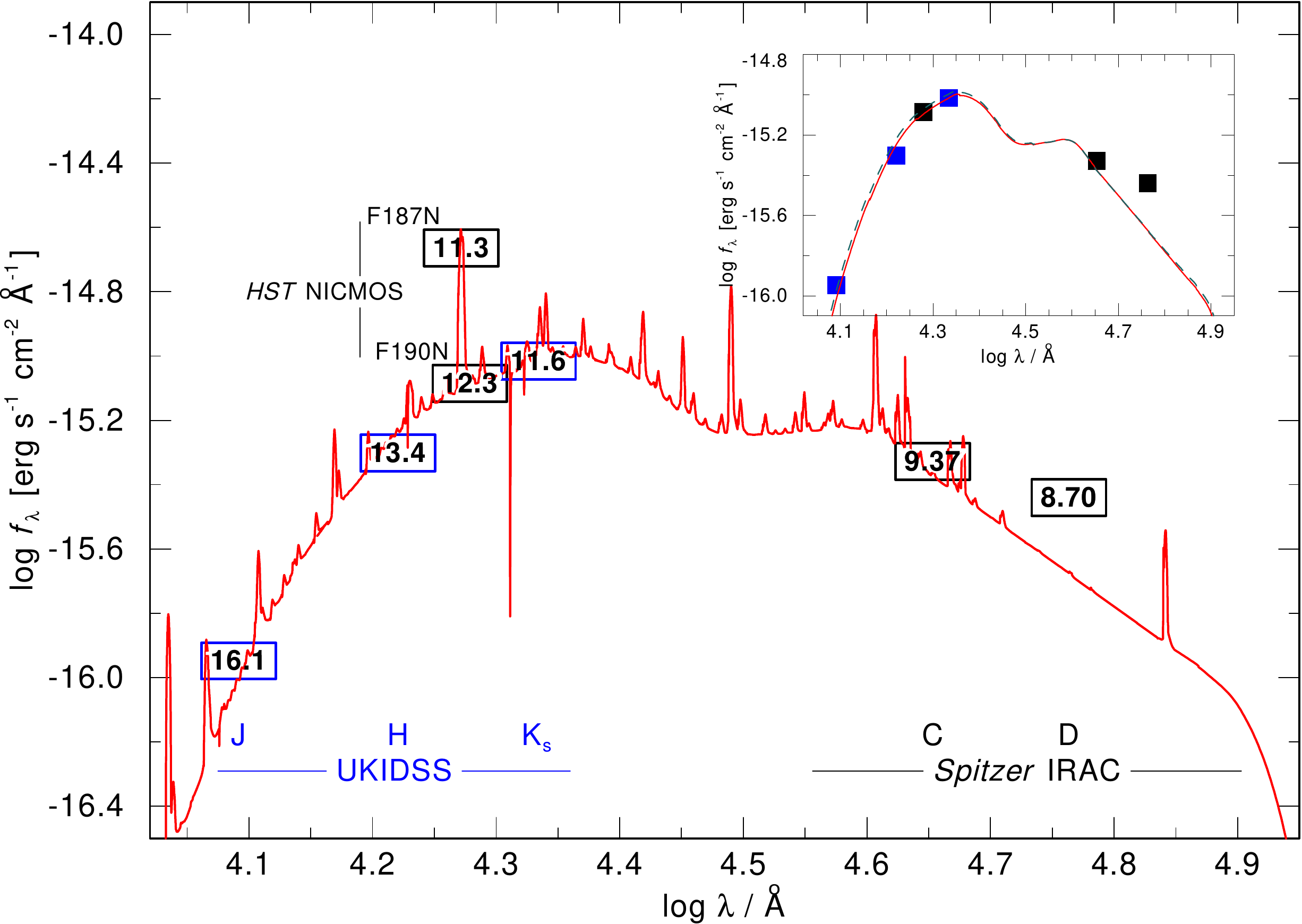}
\caption[]{
Spectral energy distribution for WR\,102c. The reddened spectrum of the PoWR
model with $T_\ast = 75\,$kK is shown as a red solid line (see Table
\ref{tab:stellarParameters102c} for the detailed parameters) and compared
to the available photometry, which is indicated by blue (UKIDSS) and black
(\textit{HST} NICMOS Pa$\alpha$ filters and \textit{Spitzer} IRAC Channels 2 and 3)
boxes.  The inset in the
upper right corner shows the continuum photometry marks
compared to the continuum SED for the two alternative models used in
Fig.\,\ref{fig:WR102c_spectrum} with $T_\ast = 66\,$kK (green
dashed curve) and $75\,$kK (red solid curve), respectively.
The overall SED fit is hardly affected by the temperature uncertainty.
We note that in the mid-IR the stellar SED is below the photometric observation, attributed
to the emission from a warm dusty nebula \citep{Barniske2008}.
}
\label{fig:WR102cSED} 
\end{figure*}

To put strong constraints on the stellar luminosity, we compared the model 
stellar spectral energy distribution (SED) with the 
available photometric and spectrophotometric data. The result is shown in 
Fig.\,\ref{fig:WR102cSED}. We used the available photometry of WR\,102c 
from 
\emph{Spitzer} IRAC \citep{Churchwell2009}, UKIDSS \citep{Lucas2008},
and \emph{HST} NICMOS \citep{Dong2011}. The SED fitting then 
allows  the luminosity and the interstellar extinction to be adjusted by adopting the  
distance modulus of $DM=14.5$\,mag. 
The observed photometry of WR\,102c is nicely reproduced by a model
with a bolometric luminosity of $\log L/L_\odot\approx5.6$ and the extinction 
$E_{B-V}=8.0$\,mag, corresponding to $A_{K}\approx2.8$\,mag. The 
alternative, slightly
cooler model ($T_\ast = 66\,$kK) leads to $\log L/L_\odot\approx5.5$ and the same
extinction. This $A_{K}$ is consistent with WR\,102c 
being located in the GC region \citep[e.g.][]{Schultheis1999,Figer1999b}.

WR\,102c is highly reddened and, if present, anomalous extinction 
may lead to an uncertainty in the estimates of stellar luminosity. To 
investigate this problem we did a thorough check of the various reddening laws 
using $R_V$ and $E_{B-V}$ provided by the extinction functions 
of \citet{Moneti2001}, \citet{Cardelli1989}, and \citet{Fitz2009}.
We found that the spectral energy distribution is best reproduced by an 
extinction of $E(B-V) = 8.0\pm 0.3$\,mag using the \citet{Moneti2001} and  
\citet{Cardelli1989} functions, while $E(B-V) = 8.7 \pm 0.4$\,mag fits best for the 
\citet{Fitz2009} function, resulting in the uncertainty in the luminosity  
$\log{L/L_\odot} = 5.5\ldots5.7$.

As a next step, we varied the $R_V$ from the commonly used value $R_V= 
3.1$ by $+0.5$ and $+1.0$ for each of the three extinction laws.
The curves were then  scaled in 
luminosity to match the {\em K}-band photometry. However, in all cases the 
reddening using the standard value $R_V= 3.1$  provides the best description of 
the stellar SED and  matches the photometry marks better.  Thus,  we 
conclude that there are no reasons to suspect an anomalous reddening for 
WR\,102c.

Our models predict that WR\,102c produces $\log(\Phi_{{\rm Ly}})= 49.0 \ldots 49.3$ 
[s$^{-1}$] hydrogen ionising photons. This compares well with the value of $49.45$
derived from radio observations of the Sickle Nebula by \citet{Lang1997}.
Thus, although our revised  number of ionising photons is lower than the  
value of \citet{Barniske2008},
WR\,102c can still be considered  the main ionising source of the Sickle nebula. 
However, the Pa$\alpha$-image of the GC region (Fig.\,\ref{fig:sickle})
does not show a fully ionised hole around WR\,102c, in contrast to the region
around the Quintuplet Cluster. This could mean that our WR star is located in
the foreground of the Sickle nebula.

\subsection{Evolutionary status of WR 102c}

\begin{figure}[ht!]
  \resizebox{\hsize}{!}{\includegraphics{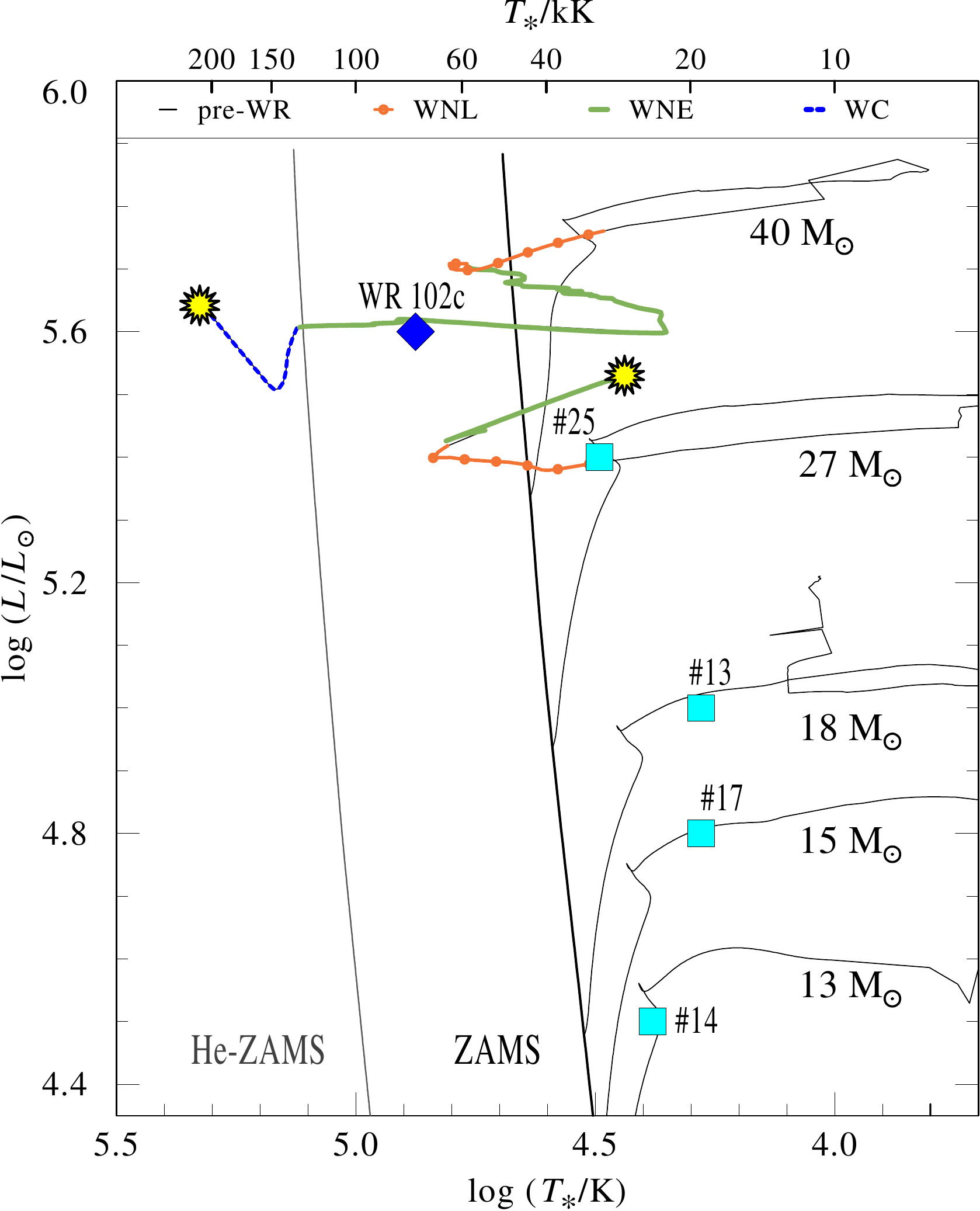}}
\caption[]{\label{fig:HRD}
 HR diagram with evolutionary tracks of \citet{Ekstroem2012} for rotating stars
of solar metallicity. The different colours identify the 
different stages of the stellar evolution: thin black: pre-WR phase; thick yellow:
WNL phase (i.e.\ with hydrogen between $40-5$\% mass fraction); thick green:
WNE phase (hydrogen less than $5$\% mass fraction); thick blue: 
WC phase (i.e.\ carbon more than $20$\% mass fraction). The positions of 
WR\,102c (blue diamond) and the OB-type stars (light blue squares) are 
indicated as well.}
\end{figure}

The derived stellar luminosity and temperature allow us to evaluate 
the evolutionary status of \mbox{WR\,102c}.  In Figure\,\ref{fig:HRD} the  position 
of WR\,102c in the HR diagram is compared with evolutionary tracks from \citet{Ekstroem2012}
for rotating stars of solar metallicity. WR\,102c is 
nicely located at the position of WNE stars on a track with an
 initial mass of $40\,M_\odot$. In the case of a rotating model, the age 
of WR\,102c is $\la 6$\,Myr, while assuming evolution without rotation 
would yield $\la 4$\,Myr. Both estimates assume single 
star evolution. With its estimated age of $4 - 6$\,Myr, WR\,102c would have 
completed at least one orbit around the GC \citep[][]{Stolte2014}. 

Assuming single star evolution and taking into 
account the ratios between numbers of  stars with different spectral types, 
\citet{Liermann2012, Liermann2014} 
constrained the age of the Quintuplet Cluster to $3\pm 0.5$\,Myr. It 
was noticed that all WN stars in the Quintuplet have late spectral types 
(WN9h) and all WC stars have WC8-9 spectral types 
\citep{Liermann2009}. The evolutionary channels that lead to the production 
of late WC stars are not yet understood \citep[see extensive 
discussion on this subject in][]{Sander2012}. The WC stars in the Quintuplet 
are severely enshrouded by dust \citep{Moneti2001}, and some of them are 
confirmed binaries \citep{Tuthill2006}.
The importance of binary channels in the evolution of massive stars is well 
recognised \citep[e.g.][]{Vanbeveren2007}. \citet{Schneider2014} used 
their binary evolution code to model the observed present-day mass functions 
of the Quintuplet Cluster and estimated its age as $4.8\pm 1.1$\,Myr if
binary channels are included.  Hence, within the uncertainties the age of 
WR\,102c is comparable to the age of the Quintuplet Cluster

WR\,102c is located $\approx 2.5$\,pc ($\approx 1^\prime$)  from the core of the 
Quintuplet Cluster. It is quite common for a massive star in the GC to 
reside outside of the three known large clusters  
\citep[e.g.][]{Cotera1999,Mauerhan2010,Dong2011}. There are different scenarios 
that can explain the origin of these isolated massive stars, for example\ they might have 
been born in relative isolation, or  ejected  or tidally stripped from one 
of the three known clusters, or they might belong to clusters that have not been  
discovered yet. 

It has recently been suggested by \citet{Lau2015} that 
the evolution of WR\,102c may have been affected by binary 
interactions, and the star could have been ejected from the Quintuplet 
Cluster. However, based on our analysis the measured radial velocity of 
WR\,102c is not outstanding and compares
well with the average radial velocity of the Quintuplet stars 
\citep[$\approx 110$\,km\,s$^{-1}$,][]{Liermann2012}. 
Proper motion observations are required to establish the runaway status of 
\mbox{WR\,102c}.   
On the other hand, \citet{Habibi2014} used their N-body simulations  
and showed that up to 80\%\ of the isolated observed WR stars in the 
GC region could be explained by tidal striping. In particular, they 
estimate the age of the Quintuplet Cluster to 5\,Myr and show that the 
projected 
tidal arms of this cluster may extend out to 60\,pc in the direction 
of the \object{Sagittarius B2} region. Given the age of WR\,102c and its proximity  
to the Quintuplet Cluster,  tidal 
stripping is able to explain the location of \mbox{WR\,102c}. 

There is, however,  another possible evolutionary path for \mbox{WR\,102c}. 
Our \textsc{SINFONI} data revealed four OB-type stars 
within a projected distance of 1 pc around WR\,102c
 (see Sect.\,\ref{sec:fstars}). Hence, it is possible that WR\,102c was formed 
together with a
its own cluster, independently from the Quintuplet Cluster (see Sect.\,\ref{sec:obcl}).

\section{Stars in the vicinity of WR 102c}
\label{sec:fstars}

\subsection{Early-type stars}

Among the 25 stellar sources detected in our \textsc{SINFONI} observations, we found 
four 
early-type stars (see Table\,\ref{tab:102c_stars}).
All of these objects have counterparts in the UKIDSS Galactic Plane 
Survey \citep{Lucas2008}; the cross-identifications are provided in 
Table\,\ref{tab:OBidentifications}.

\begin{figure}[ht!]
  \resizebox{\hsize}{!}{\includegraphics{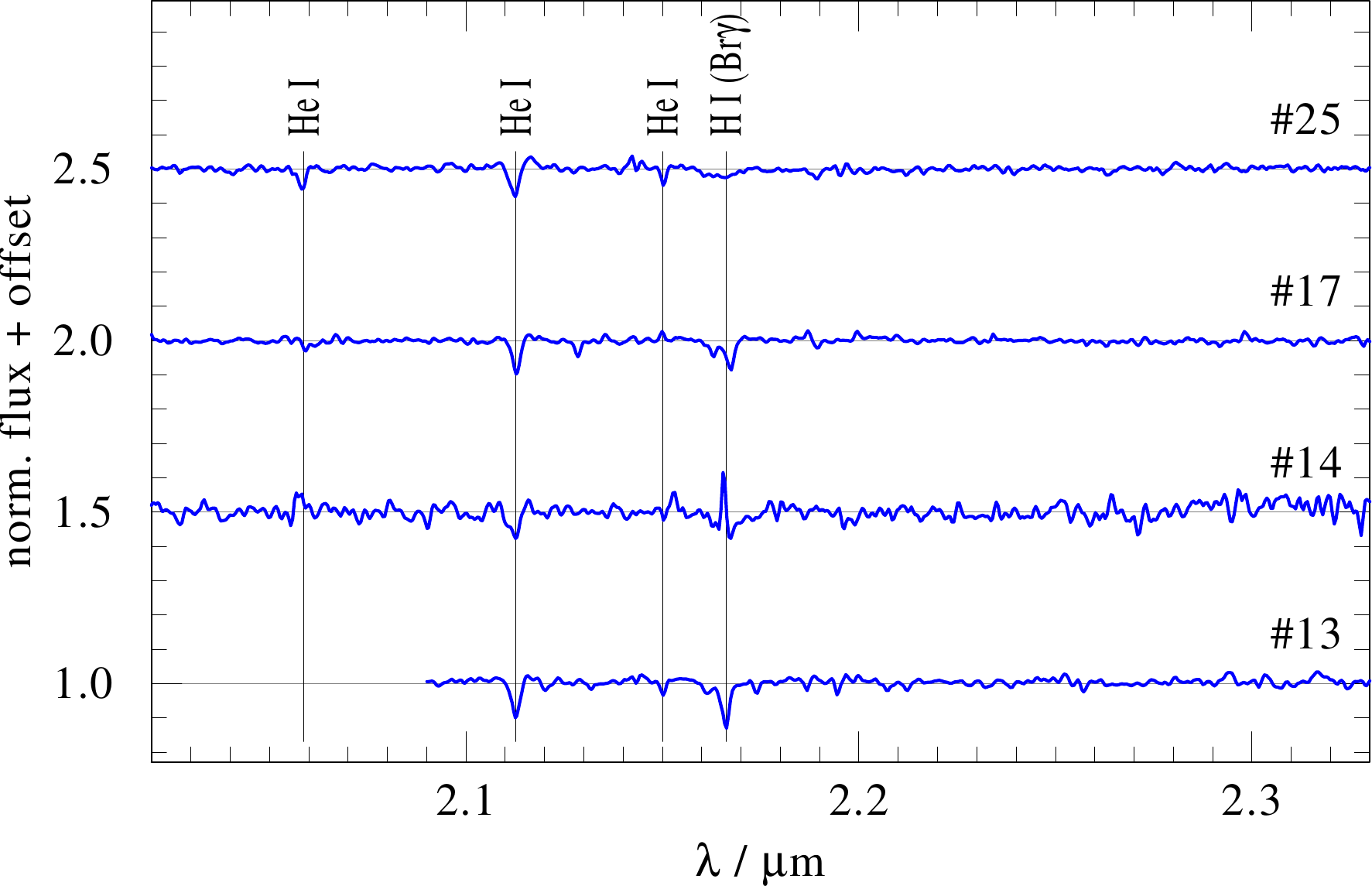}}
\caption{\label{fig:obsp}
\textsc{SINFONI} spectra of early-type stars in the WR\,102c field.
The number above each spectrum corresponds to the 
source identification in Fig.\,\ref{fig:starmap} and 
Table\,\ref{tab:102c_stars}. }
\end{figure}

The {\em K}-band spectra of these early-type stars are shown in 
Fig.\,\ref{fig:obsp}. The radial velocities measured for each star 
are consistent with WR\,102c within the uncertainties. The extinction 
to each early-type star was 
estimated from SED fitting, and was found to be similar to the 
extinction towards WR\,102c. Thus, we conclude that the 
early-type stars in the WR\,102c field are likely spatially related and 
might  even be gravitationally bound.   

For the spectral classification, we use the near-IR spectral atlas of OB 
stars \citep{Hanson1996,Hanson2005}. Source 25 is either a late O- or early B-type (super)giant. 
The other three objects are B-type stars. They hardly show   
any features in the {\em K}-band except for hydrogen Br$\gamma$ and one or two 
\ion{He}{i} lines (at $\lambda\lambda 2.06, 2.11\,\mu$m).
Moreover, Br$\gamma$ is contaminated by nebula emission. 
To overcome this contamination, we subtracted the background emission from the stellar spectra.
Therefore, it is difficult
to determine their spectral subtypes. Nevertheless, from  
the wings of Br$\gamma$ their luminosity classes could be constrained.

We have determined the stellar parameters of the OB-type stars in 
the WR\,102c cluster with tentative fits of the SED and the line
spectra to PoWR models. The results
are compiled in Table \ref{tab:stellarParameterOB}. Their initial 
masses were estimated from the comparison with evolutionary tracks in the 
the HR diagram (see Fig.\,\ref{fig:HRD}). 

\begin{table}[ht!]
\begin{center}
\caption[]{Cross-correlations in \mbox{WR\,102c} cluster with the UKIDSS 
catalogue. Source numbers refer to Table\,\ref{tab:102c_stars}}
\label{tab:OBidentifications}
\begin{tabular}{ccc}
\hline
\hline
\noalign{\vspace{1mm}}
 Source No. & & ID from UKIDSS catalogue \\
\noalign{\vspace{1mm}}
\hline
\#13  & \rule[0mm]{0mm}{3mm} & J174611.81-284906.5  \\
\#14  &  & J174611.34-284906.9 \\
\#17  &  & J174611.03-284903.9 \\
\#25  &  & J174610.87-284856.6 \\
\hline
\end{tabular}
\end{center}
\end{table}

\begin{table}[ht!]
\caption[]{Stellar parameters of the OB stars in the sample}
\label{tab:stellarParameterOB}
\centering
\begin{tabular}{lrrrr}
\hline
\hline
\noalign{\vspace{1mm}}
Source No.  & \#13   & \#14   & \#17   & \#25   \\
\noalign{\vspace{1mm}}
\hline
Spectral type        &B1-3 V: & B?III: & B2:III? & O9: I:\\
$\log L/L_\odot$ & 5.0    & 4.5    &  4.8   & 5.4    \\
$T_\ast$\, [kK]    & 20     &  24    &  20:   &31:             \\
$M_{\rm ini} [M_\odot]$
                 & 18:   & 13:    &  15   & 27:    \\
$\varv_{\rm{rad}}$ [km\,s$^{-1}$]
                 & 100    &   90   &    70  &  100   \\
$A_{K}$ [mag]    & 2.8    &  2.9    &  2.9   &  2.9   \\
\hline
  \end{tabular}
\tablefoot{
  The source number corresponds to Table\,\ref{tab:102c_stars}.
   A colon following a value flags it as uncertain. The radial velocities
   have an uncertainty of $\pm 40\,$km/s due to the limited spectral
   resolution.}
\end{table}

\subsection{Late-type stars}

In addition to the early-type stars, many point sources with late-type spectra 
are present in the field.
The temperatures were estimated from the equivalent width of the 
${}^{12}$CO (2-0) band at $2.3\,\mu$m following \citet{Gonzalez2008}. 
On the same basis we determined the spectral types. The luminosity classes 
were ranked on the basis of the CO index \citep{Blum1996,Blum2003}.
The extinction was obtained from comparison of MARCS
models \citep{Gustafsson2008}
with the measured photometric values in the infrared 
\citep{Lucas2008,Ramirez2008,Churchwell2009,Dong2011}. The distance
modulus was then inferred from this extinction by comparison with the curves 
published by \citet{Schultheis2014a}, which then also implies
the luminosity. 
The parameters of late-type stars are given in Table \ref{tab:KM-MARCS}.

\begin{table}[th!]
\begin{center}
\caption[] {The late-type stars in the field}
\label{tab:KM-MARCS}
\setlength{\tabcolsep}{4pt}
\begin{tabular}{lccccrr}
\hline
\hline
\noalign{\vspace{1mm}}
No. & $A_{K_{\rm s}}$ & $E_{B-V}$ & $d$  & $\log L$      & MK   & lum.  \\
    & [mag]           & [mag]     & [kpc]& [{$L_\odot$}] & type & class \\
\noalign{\vspace{1mm}}
\hline
 \#1&2.9&    8.4&8.0    & 2.8 & M0 &II   \\
 \#2&2.2&    6.3&5.8    & 2.1 &M1  &III  \\
 \#3&2.1&    6.0&5.5    & 1.7 &K1  & III  \\
 \#4&2.1&    6.0&5.5    & 1.5 &K4  &III  \\
 \#5&3.3&    9.5&8.1    & 2.7 &M0  &II   \\
 \#6&2.9&    8.3&8.0    & 2.7 &M0  &II   \\
 \#7&2.9&    8.3&8.0    & 2.2 &K5  &III  \\
 \#8&2.8&    8.0&8.0    & 2.3 &K4  &II   \\
 \#9&1.9&    5.6&5.3    & 1.8 &M0  &III  \\
\#10&2.9&    8.3&8.0    & 3.6 &M5-6&I    \\
\#11&2.9&    8.2&8.0    & 2.6 &K5  &II   \\
\#12&2.8&    8.0&8.0    & 3.0 &M1  &II   \\
\#16&2.6&    7.5&7.3    & 3.1 &K0  &II   \\
\#18&2.8&    8.0&8.0    & 2.3 &K3  &III  \\
\#19&2.2&    6.3&5.8    & 1.9 &K5  &III  \\
\#20&2.0&    5.7&6.2    & 1.5 &K5  &III  \\
\#21&2.5&    7.3&7.2    & 1.8 &M0  &III  \\
\#22&2.5&    7.3&7.2    & 2.1 &M0  &III  \\
\#23&2.2&    6.3&5.8    & 1.9 &K4  &III  \\
\#24&2.2&    6.3&5.8    & 1.9 &K2  &III  \\
\hline
\end{tabular}
  \end{center}   
\end{table}

\begin{figure}[ht!]
  \centering
  \resizebox{\hsize}{!}{\includegraphics{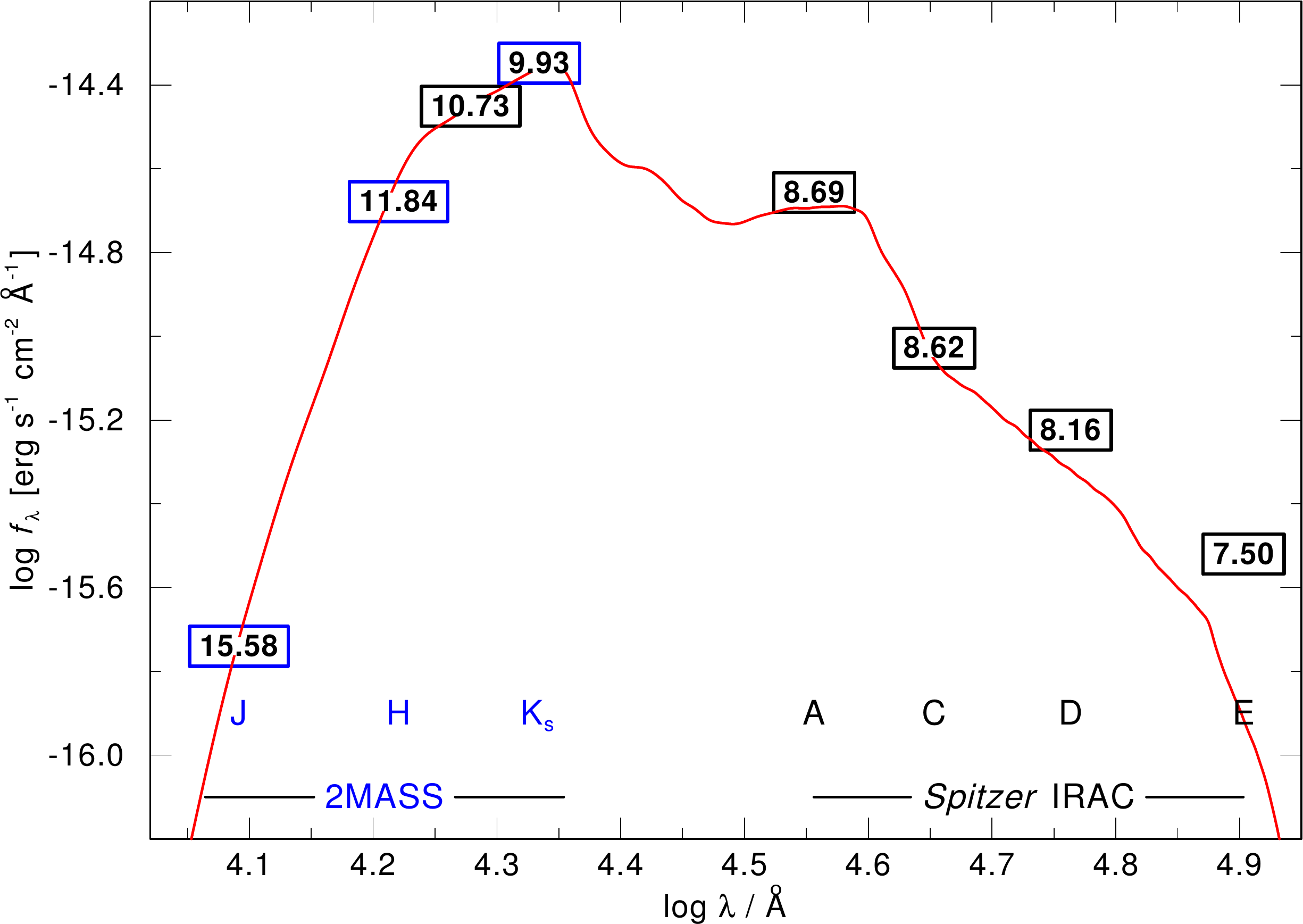}}
\caption[]{\label{fig:WR102c10sedfit}  Spectral energy distribution (SED) of 
 star 10 (\object{MSX6C G000.1668-00.0434}) 
 fitted by a MARCS model (red curve) with $T_\ast=3.3\,$kK, $\log 
L/L_\odot=3.65$, and $\log g=1.0$. The 
boxes represent photometry from 2MASS {\em J}, {\em H},\ and {\em K} 
(blue, labels = magnitudes), {\em HST} NICMOS F190N (black), and 
 {\em Spitzer} IRAC (black) in the mid-IR. The model is 
reddened with an extinction of $E_{B-V}=8.3$\,mag.}
\end{figure}

The brightest star in the field (no. 10 in Table\,\ref{tab:102c_stars} and 
Fig.\,\ref{fig:starmap}) is an M5-6I red supergiant. It is similar to the other 
late-type 
stars; we obtained its spectrum and fitted the SED as shown in 
Figure\,\ref{fig:WR102c10sedfit}. The extinction and luminosity suggest 
that the star most likely resides in the GC region. On the other 
hand, its radial velocity ($-45$\,km\,s$^{-1}$) is at odds with the radial 
velocities of early-type stars 
in this region ($\sim 100$\,km\,s$^{-1}$). The position of  star 10 in the HRD
would place it on a track for an initial mass of 9\,$M_\odot$. Such a star would 
require more than 10\,Myr to evolve to the RSG phase.

Thus, we believe that, like other late-type stars in our \textsc{SINFONI} field,  
star 10 does not belong to the \mbox{WR\,102c} cluster, but instead represents a previous 
episode of star formation in the GC region. This is similar to the conclusion 
reached by \citet{Liermann2012} from their analysis of the Quintuplet Cluster.  

\subsection{WR 102c star cluster}
\label{sec:obcl}

The presence of early-type stars within 0.9\,pc of WR\,102c   may indicate that 
these objects represent an independent cluster. 
Alternatively, the \mbox{WR\,102c} cluster may be related to the same starburst that 
produced the Quintuplet Cluster. Yet another possibility is that all the early-type 
stars close to WR\,102c are part of the Quintuplet's tidal tail. 
Such tails have been predicted by \citet{Habibi2014} for the \object{Arches} and Quintuplet
with extensions up to $1.6$\,pc from the cluster centre.
In any case, the presence of a cluster around WR\,102c seems to be 
inconsistent with the scenario that WR\,102c was born in the Quintuplet and ejected 
from there. 

We used the publicly available cluster evolution code {\sc McLuster} 
\citep{Kuepper2011} to estimate the mass of the \mbox{WR\,102c} cluster. Assuming 
a standard Kroupa IMF \citep{Kroupa2001}, the mass of the \mbox{WR\,102c} cluster is $\sim 
1000\,M_\odot$. The simulation accurately reproduces the mass of the most 
massive 
star and the number of early-type stars we observed.  Given the sensitivity 
of our observations ($K_{\rm s}=14-15$\,mag) we were able to detect only stars 
earlier than B1-2, i.e.\ with masses above $10\,M_\odot$.

Adopting an enclosed mass of $\sim 1000\,M_\odot$ and a characteristic cluster 
radius $R_0\approx 1$\,pc we estimate the velocity dispersion within the 
cluster stars to be only a few km\,s$^{-1}$ \citep[following equation\,1 
in][]{Kroupa2002}. 
In summary, the number of early-type stars, their ages, masses, and radial 
velocities consistently agree with these objects being the brightest members 
of a yet unknown star cluster.

\section{Summary and conclusions}
\label{sec:sum}
Using  VLT \textsc{SINFONI} we obtained integral field observations of 0.25
square arcmin area 
around  WR\,102c located at $\approx 2.5$\,pc (projected distance)  from 
the Quintuplet Cluster in the GC region. 

We identified five early-type stars  (including 
\mbox{WR\,102c}) in this field. Based on their age, radial velocity, and mass distribution 
we suggest that these are the most massive members of a bound \mbox{WR\,102c} cluster 
with $M_{\rm cl} \sim 1000\,M_\odot$.  The existence of such a cluster 
questions the ejection scenario for the present day location of \mbox{WR\,102c}.  

The detailed analysis of the stellar spectrum and photometry of WR\,102c
by means of non-LTE stellar atmosphere PoWR models yielded stellar and wind 
parameters typical for a WN6 star (Table\,\ref{tab:stellarParameters102c}).
The uncertainty of the deduced stellar parameters of WR\,102c is mild 
and might reflect some departures of stellar wind from spherical 
symmetry, e.g.\ due to the fast rotation or binarity.
  
We produced a map of diffuse Br$\gamma$ emission in the region around \mbox{WR\,102c}. 
Our data clearly show a bipolar or ring-shaped nebula centred on \mbox{WR\,102c}. 
The same nebula is clearly seen in the images from the {\em HST} Paschen-$\alpha$ survey 
of the GC. According to the spectra of the nebula, the projected 
expansion velocity is small.  We suggest that the
\mbox{WR\,102c} nebula is a typical WR nebula formed recently as the result 
of the interaction of winds at different evolutionary stages.

\section*{Acknowledgments}   
This work has extensively used the NASA/IPAC Infrared Science Archive, 
the NASA Astrophysics Data System, and the SIMBAD database, operated at 
CDS, Strasbourg, France. This publication makes use of data products 
from the Two Micron All Sky Survey, which is a joint project of the University 
of Massachusetts and the Infrared Processing and Analysis Center/California 
Institute of Technology, funded by the National Aeronautics and Space
Administration and the National Science Foundation. 
Some of the data presented in this paper were retrieved from the Mikulski 
Archive for Space Telescopes (MAST). STScI is operated by the Association of 
Universities for Research in Astronomy, Inc., under NASA contract NAS5-26555. 
Support for MAST for non-{\em HST} data is provided by the NASA Office of Space 
Science via grant NNX09AF08G and by other grants and contracts. We are grateful 
to Dr. R. Lau for sharing with us the preprint of his paper. 
We thank the referee for the useful comments and suggestions.
Funding for this research has been provided by DLR grant 50\,OR\,1302 (LMO) and 
DFG grant HA\,1455/26 (AS). 

\bibliographystyle{aa}
\bibliography{wr102c}

\end{document}